\documentclass{elsart5p}

\usepackage{graphicx}
\usepackage{amssymb,amsmath}

\newcommand{\Eq}{Eq.}

\newcommand{\Fig}{Fig.}

\newcommand{\Ref}{Ref.}
\newcommand{\Refs}{Refs.}
\newcommand{\Sect}{Section}

\newcommand {\mct}{\mathcal{T}}

\newcommand{\He}{{}^3\mathrm{He}}
\newcommand{\HE}{{}^4\mathrm{He}}
\newcommand{\Hh}{{}^3\mathrm{H}}
\newcommand{\nH}{n\text{-}{}^3\mathrm{H}}
\newcommand{\pHe}{p\text{-}{}^3\mathrm{He}}
\newcommand{\pH}{p\text{-}{}^3\mathrm{H}}
\newcommand{\nHe}{n\text{-}{}^3\mathrm{He}}

\newcommand{\dd}{d\text{-}d}

\begin{document}

\begin{frontmatter}
\title {Four-nucleon system with $\Delta$-isobar excitation}
 
\author[cfnul]{A.~Deltuva},
\ead{deltuva@cii.fc.ul.pt}
\author[cfnul]{A.~C.~Fonseca}, 
\author[itpuh]{P.~U.~Sauer}
\address[cfnul]{Centro de F\'{\i}sica Nuclear da Universidade de Lisboa, 
P-1649-003 Lisboa, Portugal }
\address[itpuh]{Institut f\"ur Theoretische Physik, Leibniz Universit\"at Hannover,
  D-30167 Hannover, Germany}


\begin{abstract}
The four-nucleon bound state and scattering below three-body breakup
threshold are described based on the realistic
coupled-channel potential CD Bonn + $\Delta$ which allows the excitation
of a single nucleon to a $\Delta$ isobar. The Coulomb repulsion between
protons is included. In the four-nucleon system the two-baryon 
coupled-channel potential yields effective two-, three- and four-nucleon forces,
mediated by the $\Delta$ isobar and consistent with each other and
with the underlying two-nucleon force.
The effect of the four-nucleon force on the studied observables is 
much smaller than the effect of the three-nucleon force.
The inclusion of the $\Delta$ isobar is unable to resolve the existing
discrepancies with the experimental data.
\end{abstract}


\end{frontmatter}

\section{Introduction \label{sec:intro}}

State of art calculations of four-nucleon $(4N)$ scattering have been
recently presented in \Refs~\cite{deltuva:07a,deltuva:07b,deltuva:07c}
for all possible reactions initiated by $\nH$, $\pHe$, $\nHe$, $\pH$ and
$\dd$ below three-body breakup threshold. Realistic two-nucleon $(2N)$ 
interactions based on meson theory like AV18~\cite{wiringa:95a}, 
CD Bonn~\cite{machleidt:01a} and INOY04~\cite{doleschall:04a} or chiral
effective field theory (EFT) \cite{entem:03a} are used between pairs together 
with the Coulomb repulsion
between the protons. No approximations were used in the solution of
the four-body scattering equations beyond the usual partial-wave 
decomposition and the discretization of integration variables.
The results presented are fully converged vis-a-vis the included partial
waves as well as the number of mesh points used for the discretization of
all continuous variables.
Some observables we obtain are described quite well by all 
interaction models, some scale with the three-nucleon $(3N)$ binding energy, 
and some show large deviations from the data.

Therefore the next step in our understanding of $4N$  observables in terms
of the underlying forces between nucleons requires the inclusion of a
$3N$ force. There are three distinct ways for doing this: 
a) Add a static two-pion-exchange 
irreducible $3N$ force \cite{coelho:83a,coon:79a,pudliner:97a} 
to the underlying $2N$ forces;
 however, in this approach these two forces are not really consistent with each other. 
b) Use $2N+3N$ force models based on chiral EFT \cite{kolck:99a,epelbaum:00a}
to guaranty  consistency between the $2N$ and $3N$ forces; however, for a 
realistic description, the expansion up to at least next-to-next-to-next-to
leading order (N3LO) is required, which for the $3N$ force is not yet available.
c) Extend the purely nucleonic model to allow the explicit excitation
of a nucleon $(N)$ to a $\Delta$ isobar as was carried out 
in \Ref~\cite{deltuva:03c} for the $2N$ and $3N$ systems; this approach 
yields effective many-nucleon forces consistent with the underlying $2N$ force, 
but does not fully satisfy chiral symmetry, much like a).

The studies of the $3N$ system reveal that all these different approaches
lead to qualitatively similar results. In the $4N$ system the first one a)
was already applied to $\nH$ \cite{lazauskas:04a,lazauskas:05a} and
$\pHe$ \cite{viviani:01a,fisher:06} scattering. 
In the present paper, following the work on the $3N$ system performed in 
\Refs\cite{deltuva:03c,deltuva:05a,deltuva:05d}, 
 we use the last approach c) to study all $4N$ reactions below three-body
breakup threshold. In the $3N$ system the excitation of a single nucleon 
to a $\Delta$ isobar yields an effective $3N$ force with components of
Fujita-Miyazawa type \cite{fujita:57a} and much richer structures
in a reducible form; beside pion $(\pi)$ exchange, the $3N$ force
has contributions of shorter range due to the exchange of heavier mesons.
In the $4N$ system an effective $4N$ force arises that also has parts of
shorter range than $\pi$ exchange.

The paper introduces the dynamics chosen for the extended description
of the $4N$ system in \Sect~\ref{sec:dyn}.
It discusses the effects of $\Delta$-isobar excitation in the form of $3N$ 
and $4N$ forces on the $4N$ bound state in \Sect~\ref{sec:bs} 
and on low-energy $4N$ scattering observables  in \Sect~\ref{sec:scatt}.
Conclusions are given in \Sect~\ref{sec:concl}.

\section{Dynamics \label{sec:dyn}}

\begin{figure}[b]
\begin{center}
\includegraphics[scale=0.32]{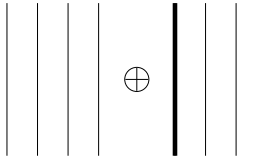}
\end{center}
\caption{ \label{fig:hs} 
Four-baryon Hilbert space considered. It consists of a purely nucleonic sector
$\mathcal{H}_N$ and a sector $\mathcal{H}_\Delta$ in which one nucleon
is turned into a $\Delta$ isobar, indicated by a thick line.}
\end{figure}

The description of the $4N$ system is given in a Hilbert space consisting of 
two sectors as depicted in \Fig~\ref{fig:hs}; the first sector $\mathcal{H}_N$
is purely nucleonic, and in the second sector $\mathcal{H}_\Delta$ one nucleon 
is replaced by a $\Delta$ isobar of mass
$m_\Delta = 1232$ MeV. The restriction to Hilbert sectors with one
$\Delta$ at most has a strong physics motivation. The single $\Delta$ isobar,
when coupled to explicit pion-nucleon states, mediates the $P_{33}$
resonance in pion-nucleon scattering; it also mediates single-pion
production in $2N$ scattering where single-pion production is the 
dominant inelastic channel up to about 500 MeV in the $2N$ center of mass
(c.m.) system, i.e., far beyond the two-pion threshold.
Thus, the adopted Hilbert space is sufficient for a further extension
to the intermediate-energy pionic channels where a Hilbert sector with two 
pions appears to be dynamically suppressed.
The nucleons in the sector $\mathcal{H}_N$ are fully antisymmetrized.
The sector $\mathcal{H}_\Delta$ does not has a physics life on its own, 
but is included only through its coupling to $\mathcal{H}_N$. That coupling 
is symmetric in all nucleons. Though the $\Delta$ isobar is physically
distinct from the nucleons, only wave-function components, antisymmetrized
in all four baryons, nucleons and the $\Delta$ isobar, have to be 
considered. Thus,  Faddeev-Yakubovsky bound-state equations in the 
symmetrized form of \Ref~\cite{kamada:92a} 
and Alt, Grassberger and Sandhas (AGS) scattering equations  
 \cite{grassberger:67} in the symmetrized form of 
\Refs~\cite{deltuva:07a,deltuva:07b,deltuva:07c} can be used.

\begin{figure}[!]
\begin{center}
\includegraphics[scale=0.36]{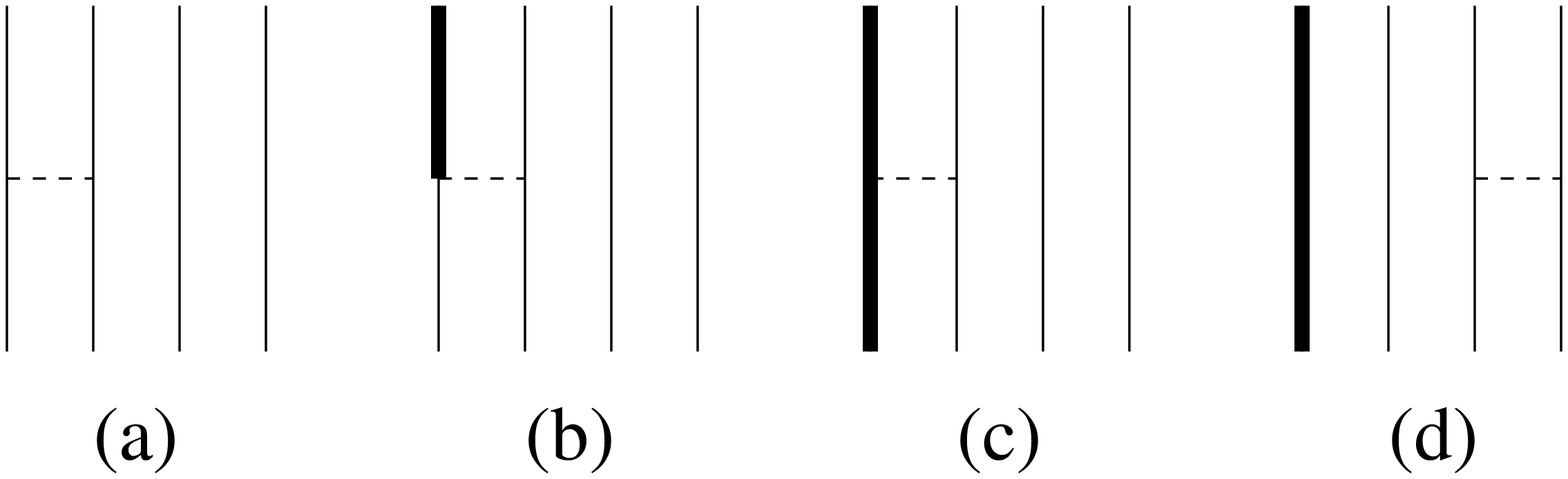}
\end{center}
\caption{ \label{fig:H} 
Four-baryon Hamiltonian. The dashed horizontal lines indicate potentials.}
\end{figure}

The dynamics is specified by a hermitian Hamiltonian $H$ with
instantaneous two-baryon potentials as indicated in \Fig~\ref{fig:H}
for the $4N$ system.
The Hamiltonian acts in both Hilbert sectors $\mathcal{H}_N$ and
$\mathcal{H}_\Delta$ and couples them. The hermitian-conjugate of
the component (b) is not shown separately.
When limited to the $2N$ system, the Hamiltonian of \Fig~\ref{fig:H}
 (a) - (c), i.e., its respective components  $v_{NN}$, 
$v_{\Delta N} = v_{N \Delta}^{\dagger}$ and $v_{\Delta \Delta}$,
reduces to the potential CD Bonn + $\Delta$, a realistic
coupled-channel two-baryon potential, fitted in \Ref~\cite{deltuva:03c}
to the elastic $2N$ data. The Hamiltonian component of \Fig~\ref{fig:H} (d),
corresponding to the $2N$ potential in the presence of a $\Delta$ isobar, is not 
constrained by $2N$ data. A reasonable choice is the purely nucleonic CD Bonn 
potential \cite{machleidt:01a} which we used in our previous $3N$ calculations.
However, we found that the results for $3N$ observables depend extremely
weakly on the parametrization of the potential in \Fig~\ref{fig:H} (d).
Even choosing it to be zero has no visible consequences on the description
of $3N$ observables; e.g., the calculated $3N$ binding energy changes by
20 keV only. Therefore  we choose the Hamiltonian component of 
\Fig~\ref{fig:H} (d) to be zero in our $4N$ calculations. That choice is an 
assumption on unknown dynamics, but it also yields a technical simplification. 
The solution of the $4N$ equations remains exact.

\subsection {Equations \label{sec:eq}}

The symmetrized equations for the Faddeev-Yakubovsky amplitudes 
$|\psi_\alpha \rangle$ of the $4N$ bound state are
\begin{subequations} \label{eq:fy}
\begin{align}  \label{eq:fy1}
|\psi_1 \rangle = {}&  G_0 T G_0 U_1(-P_{34}|\psi_1 \rangle + |\psi_2 \rangle), 
\\ \label{eq:fy2}
|\psi_2 \rangle  = {}&  G_0 T G_0 U_2 (1-P_{34})|\psi_1 \rangle, 
\end{align}
\end{subequations}
where $G_0$ is the free four-particle Green's function and $T$ the
two-baryon transition matrix.
The  operators $U_\alpha$ obtained from
\begin{subequations}
\begin{align}
\label{eq:U}
U_{\alpha} = {} & P_\alpha G_0^{-1} +
P_\alpha \, T\, G_0 \, U_{\alpha}, \\
\label{eq:P}
P_1 = {} & P_{12}\, P_{23} + P_{13}\, P_{23}, \\
\label{eq:tildeP}
P_2 = {} & P_{13}\, P_{24},
\end{align}
\end{subequations}
are the symmetrized AGS operators for the $1+3$ and $2+2$ subsystems
and $P_{ij}$ is the permutation operator of particles $i$ and $j$.
The equations suffice for calculating the binding energy.
The step from the Faddeev-Yakubovsky amplitudes $|\psi_\alpha \rangle$
to the bound state wave function is not yet carried out.

The corresponding equations for $4N$ scattering  and the description of the
screening and renormalization method to include the Coulomb interaction
are given in \Refs~\cite{deltuva:07a,deltuva:07b,deltuva:07c}, and
for that reason are not repeated here.

\subsection{The isolation of $\Delta$-isobar effects \label{sec:Deff}}

\begin{figure}[!]
\begin{center}
\includegraphics[scale=0.36]{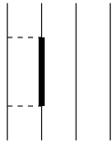}
\end{center}
\caption{ \label{fig:2Ndisp} 
Lowest order contribution to the $2N$ dispersion.}
\end{figure}

Full four-body calculations are carried out in \Sect~\ref{sec:bs} for the
$4N$ bound state and in \Sect~\ref{sec:scatt} for selected $4N$ reactions.
The dynamics is based on the coupled-channel potential CD Bonn + $\Delta$;
the purely nucleonic CD Bonn potential serves as reference for isolating
the full $\Delta$-isobar effect on the considered observables.
However, a split of the total $\Delta$-isobar effect into separate
contributions is highly desirable for understanding the physics of the
results. For this goal a sequence of
incomplete calculations is also done. The dynamic input,  the 
coupled-channel two-baryon transition matrix is calculated correctly
in all its components $T_{NN}$, $T_{\Delta N}$, $T_{N \Delta}$, and
$T_{\Delta \Delta}$, but is only partially included in the following 
incomplete calculations:

(1) Only the purely nucleonic component $T_{NN}$ of the two-baryon 
transition matrix is retained. The lowest order $\Delta$ contribution to the
dynamics, kept in this calculation, is shown in Fig.~\ref{fig:2Ndisp}.
It renders the $2N$ interaction less attractive off-shell.
This is the well known effect of $2N$ dispersion.

\begin{figure}[!]
\begin{center}
\includegraphics[scale=0.36]{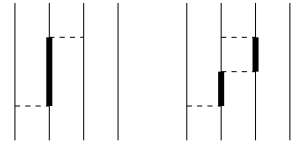}
\end{center}
\caption{ \label{fig:FM} 
Effective $3N$ force of the Fujita-Miyazawa type (left side) and an example 
for a higher order $3N$ force (right side) that is included together.}
\end{figure}

(2) Only the two-baryon transition matrix components $T_{NN}$, $T_{\Delta N}$, 
and $T_{N \Delta}$ are retained. The most important $\Delta$ contribution to the
dynamics, kept in this calculation in addition to the $2N$ dispersion, is
of Fujita-Miyazawa (FM) type  shown in Fig.~\ref{fig:FM} together
with higher order $3N$ force contributions; the sample process on the
right-hand side of Fig.~\ref{fig:FM} occurs due to two-body contributions
contained in $T_{N \Delta}$. However, according to \Ref~\cite{deltuva:03b}, 
those higher order $3N$ force contributions should be far less important.
Thus, the second incomplete calculation, when compared to the first one,
is a reasonable estimation for the effective $3N$ force of the
Fujita-Miyazawa type.

\begin{figure}[!]
\begin{center}
\includegraphics[scale=0.36]{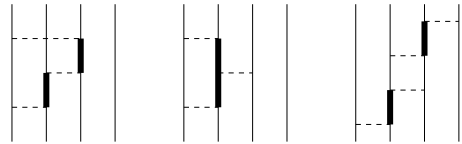}
\end{center}
\caption{ \label{fig:3nfho} 
Examples for higher order $3N$ force processes. The first two diagrams show 
contributions to the $3N$ force mediated by the two-baryon transition matrix 
component $T_{\Delta \Delta}$ contained in $U_1$; one nucleon stays uninvolved.
 In the last diagram all four nucleons interact; the process is the iteration 
of the $3N$ Fujita-Miyazawa force; it is due to the purely nucleonic 
intermediate states in $T_{\Delta \Delta}$.}
\end{figure}

\begin{figure}[!]
\begin{center}
\includegraphics[scale=0.36]{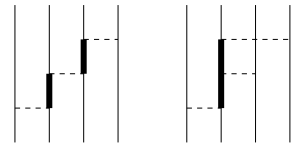}
\end{center}
\caption{ \label{fig:4nf} 
Lowest order effective $4N$ force contributions.}
\end{figure}

(3) In the third incomplete calculation all $4N$ force effects are attempted
to be eliminated while keeping all $3N$ force effects. 
The two-baryon transition matrix component $T_{\Delta \Delta}$, contained in
the $1+3$ subsystem transition operator $U_1$, generates higher 
order (h.o.) $3N$ force contributions like those in Fig.~\ref{fig:3nfho}, 
in which one spectating nucleon is interaction-free. In addition, 
due to the purely nucleonic intermediate states in $T_{\Delta \Delta}$, 
even particular iterations of the $3N$ Fujita-Miyazawa force are generated
like the one also shown in Fig.~\ref{fig:3nfho}, 
in which all four baryons are involved in the interaction process. 
But $T_{\Delta \Delta}$ is also the source for the effective $4N$ force, whose
corresponding lowest order contributions are shown in Fig.~\ref{fig:4nf}.
The clean elimination of the $4N$ force is achieved by using the full
$T_{\Delta \Delta}$ component in the calculation of $U_1$, but the modified part 
$T_{\Delta \Delta}- T'_{\Delta \Delta}$ when the transition matrix acts immediately 
before/after the permutation operator $P_{34}$ in the iteration process of 
\Eq~\eqref{eq:fy1} where the particular transition matrix to be modified occurs 
explicitly. The subtraction of 
$T'_{\Delta \Delta} = v_{\Delta \Delta} (1+G_0 T'_{\Delta \Delta})$ ensures the 
presence of the purely nucleonic intermediate state between two successive $3N$ 
transition operators $U_1$, which act, due to the permutation $P_{34}$, in 
different $3N$ subsystems. It therefore eliminates all $4N$ force contributions
for which  Fig.~\ref{fig:4nf} gives lowest order examples.
Thus, when comparing this calculation to the previous incomplete calculation (2)
and to the full calculation, the effects of h.~o. $3N$ force 
contributions and of the $4N$ force are estimated separately.

\section{Four-nucleon bound state \label{sec:bs}}

In the present study the binding energy for the $4N$ bound state is 
calculated. Other properties, such as the charge radius or the charge
form factor, are not determined yet. Only total isospin
$\mct = 0$ states are included and isospin averaging is performed 
for the two-baryon transition matrix. 

In Table \ref{tab:eb} we collect the results for  $3N$ and $4N$ binding 
energies. The inclusion of the $\Delta$ isobar
increases the corresponding binding energies but is unable to reproduce
the experimental values. Obviously, many-nucleon forces,
not accounted for by the $\Delta$ isobar, make a rather significant
contribution to $3N$ and $4N$ binding energies.
 Table \ref{tab:eb} also splits up the total  $\Delta$-isobar effect
into separate contributions obtained through incomplete calculations 
as discussed in \Sect~\ref{sec:Deff}.
(1) The $2N$ dispersion turns out to be massive in the $4N$ bound state with
$\Delta E_2 = -2.80$ MeV; it arises mainly from the dispersion in the 
${}^1S_0$ $2N$ partial wave. (2) The $3N$ force contribution of the
Fujita-Miyazawa type $\Delta E_3^{\mathrm{FM}} = 2.25$ MeV is also quite large.
The increase by the factor of 4.5 compared to the $3N$ bound state
is understandable in terms of the different multiplicity with which the
$3N$ force contributes: one in the $3N$ bound state and four in the $4N$.
The observed factor of $\approx 4.5$ comes from the fact that $\HE$, 
being a denser system 
than $\He$ or $\Hh$, squeezes out more binding from the underlying force.
(3) The contribution of the higher order $3N$ force terms mediated by the 
diagonal $N\Delta$ potential $v_{\Delta \Delta}$ is
$\Delta E_3^{\mathrm{h.o.}} = 1.30$ MeV comparable to the one of 
the Fujita-Miyazawa type. The size of these h.~o. terms depends on the strength
of the $\sigma$-meson exchange in $v_{\Delta \Delta}$ that is not
really constrained by elastic $2N$ data; it could get constrained by
the data in pionic channels coupling to $2N$ channels above the inelastic
threshold. The $\sigma$-meson exchange strength of the CD Bonn + $\Delta$
potential was chosen to yield more binding in the $3N$ system,
and therefore its contribution to the binding energy of $\HE$ is
quite sizable as well.
An alternative realistic coupled-channel potential with a weaker 
$\sigma$-meson in $v_{\Delta \Delta}$ was developed in \Ref~\cite{deltuva:04b}.
In the Appendix we show the differences relative to CD Bonn + $\Delta$ and 
the corresponding predictions for $3N$ and $4N$ binding energies.
(4) Finally, in contrast to the complete $3N$ force contribution
$\Delta E_3^{\mathrm{FM}} + \Delta E_3^{\mathrm{h.o.}} = 3.55$ MeV, 
the contribution arising from the effective $4N$ force 
$\Delta E_4 = 0.17$ MeV is indeed rather small.

\begin{table}[!]
\begin{tabular}{l*{3}{@{\quad\quad}r}}
\hline
& ${}^3\mathrm{H}$ & ${}^3\mathrm{He}$ & ${}^4\mathrm{He}$ \\ \hline
CD Bonn  & 8.00 & 7.26 & 26.18  \\
CD Bonn + $\Delta$ &  8.28 & 7.53 &   27.10 \\
exp &   8.48 & 7.72 & 28.30  \\ \hline
 $\Delta E_2$ & -0.51 & -0.48  & -2.80  \\
$\Delta E_{3}^{\mathrm{FM}}$ & 0.50 & 0.48 & 2.25 \\
$\Delta E_3^{\mathrm{h.o.}}$ & 0.29 & 0.27 & 1.30 \\
$\Delta E_{4}$ & & & 0.17 \\
\hline
\end{tabular}
\caption{ \label{tab:eb}
Binding energies for $\Hh$, $\He$, and $\HE$ derived from the potentials
CD Bonn and CD Bonn + $\Delta$ and the corresponding experimental values 
are  given in the first three rows.
The last four rows split the complete $\Delta$ effect up into
$2N$ dispersion $\Delta E_2$, 
Fujita-Miyazawa type $3N$ force effect $\Delta E_{3}^{\mathrm{FM}}$, 
higher order $3N$ force effect $\Delta E_3^{\mathrm{h.o.}}$, and
$4N$ force effect $\Delta E_{4}$ for $\HE$.
All results are given in MeV.}
\end{table}

\section{Four-nucleon scattering \label{sec:scatt}}

The $\nH$ and $\pHe$ scattering is 
dominated by the total isospin $\mct = 1$ states while deuteron-deuteron 
$(\dd)$ scattering by the $\mct = 0$ states; the $\nHe$ and $\pH$ reactions 
involve both $\mct = 0$ and $\mct = 1$ states and are coupled to $\dd$ in 
$\mct = 0$. All those reactions below three-body breakup threshold were 
calculated in \Refs~\cite{deltuva:07a,deltuva:07b,deltuva:07c}.
Here we study the $\Delta$-isobar effect on the low-energy $4N$ scattering
 observables.

\begin{figure}[!]
\begin{center}
\includegraphics[scale=0.65]{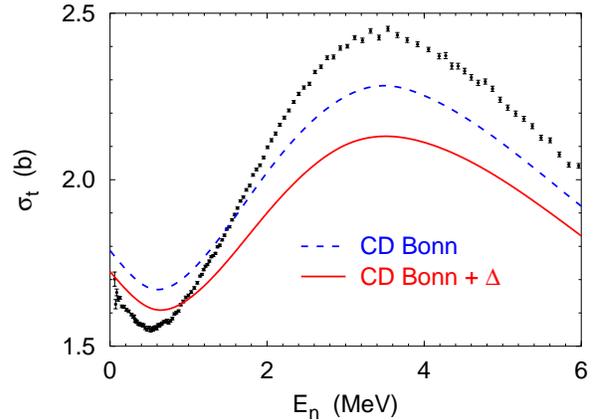}
\end{center} 
\caption{ \label{fig:nt} 
Total cross section for $\nH$ scattering as function of neutron lab energy 
calculated with the CD Bonn (dashed curve) and CD Bonn + $\Delta$ (solid curve)
potentials. Experimental data are from \Ref~\cite{phillips:80}.}
\end{figure}

In Fig.~\ref{fig:nt} we study the energy dependence of the total $\nH$ cross 
section. The $\Delta$-isobar excitation increases the 
$3N$ binding energy and through scaling improves the description of the
data around threshold. However, there is quite a significant non-beneficial
$\Delta$ effect in the region of the resonance which is strongly driven
by $\nH$ relative $P$ waves. In Table \ref{tab:nt} we split this effect into
$2N$ dispersion, $3N$ and $4N$ force contributions,  the last of which
we find to be negligible. Furthermore, we split the total $\nH$ cross section 
$\sigma_t$ into the $S$- and $P$-wave contributions $\sigma_t^S$ and
$\sigma_t^P$. As can be seen in Table \ref{tab:nt},
the $\Delta$ effects on the $\Hh$ binding energy $\epsilon_t$ and the
$S$-wave cross section are correlated by scaling in the same way as it has
been observed in \Ref~\cite{deltuva:07a}, i.e., $\sigma_t^S$ decreases
when $\epsilon_t$ increases. In contrast,
there is no such a correlation in $P$ waves  where both $2N$ dispersion and 
effective $3N$ force decrease the cross section while having opposite effects
on $\epsilon_t$.
This is different from $3N$ scattering where the $\Delta$ effect becomes
visible, scaling aside, only at rather high energy,
beyond 50 MeV in the center of mass (c.m.) system. At lower energies the individual
$\Delta$ contributions are not negligible, but very often cancel each
other to a large extent. Much smaller effect on $\sigma_t$ in the resonance region
is observed in \Refs~\cite{lazauskas:04a,lazauskas:05a} where Urbana IX $3N$ force 
is added to AV18.

\begin{table}[!]
\begin{tabular}{l*{5}{@{\quad}r}}
\hline
& $\epsilon_t$ & $\sigma_t^S$ & $\sigma_t^P$ & $\sigma_t$ &
$\mathrm{A_y^{max}}$  \\ \hline
CD Bonn  & 8.00 & 0.975 & 1.308 & 2.283 &   0.364 \\
CD Bonn + $\Delta$ &  8.28 & 0.958 & 1.172 & 2.130 &  0.345 \\
exp &  8.48  &  &  & 2.450 &  \\ \hline
$2N$ dispersion & -0.51 & 0.036 & -0.075 & -0.039 &  -0.055\\
$3N$ force (FM) & 0.50 & -0.035 & -0.058 & -0.094 & 0.022 \\
$3N$ force (h.o.) & 0.29 & -0.017 & -0.004 & -0.021 & 0.014 \\
$4N$ force & & $<0.001$ & $<0.001$ &  $<0.001$ & $<0.001$ \\
\hline
\end{tabular}
\caption{ \label{tab:nt}
Separate $\Delta$-isobar effects on the observables of $\nH$ scattering 
at 3.5 MeV neutron lab energy.}
\end{table}

\begin{figure}[!]
\begin{center}
\includegraphics[scale=0.5]{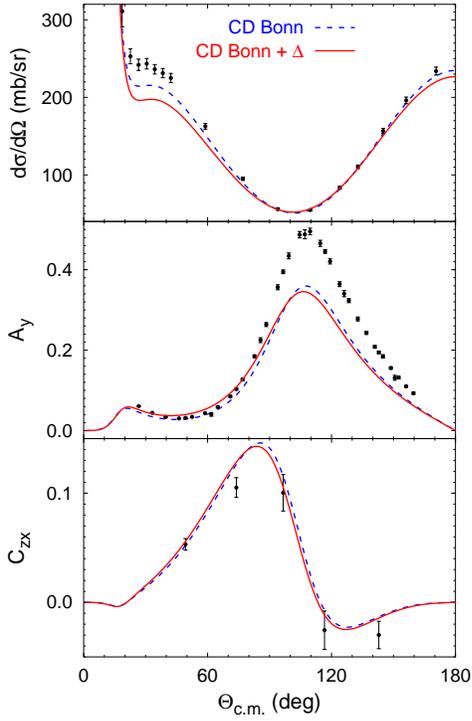}
\end{center}
\caption{\label{fig:p3He} 
Differential cross section, proton analyzing power $A_{y}$ and $\pHe$ 
spin correlation coefficient $C_{xz}$ of $\pHe$ scattering at 5.54 MeV 
proton lab energy as functions of c.m. scattering angle.
Curves as in \Fig~\ref{fig:nt}.
The data are from \Ref~\cite{mcdonald:64} for the differential cross section 
and from \Ref~\cite{alley:93} for the spin observables. }
\end{figure}

In Fig.~\ref{fig:p3He} we study the observables of $\pHe$ scattering
at 5.54 MeV proton lab energy. This reaction is related to $\nH$
by charge symmetry that is broken only by the Coulomb interaction and 
hadronic charge dependence. The charge-symmetric
$\Delta$ effect is therefore very similar in both reactions. 
It reduces the $\pHe$ differential cross section at forward and
backward angles increasing the discrepancy with data.
It is small and non-beneficial for the proton analyzing power $A_y$,
while the $\pHe$ spin correlation coefficients remain described quite
satisfactorily. 
Similar effects have also been observed in \Ref~\cite{fisher:06}
using the Urbana IX $3N$ force \cite{pudliner:97a}.
In the last column of Table \ref{tab:nt} we split the $\Delta$ effect into 
$2N$, $3N$ and $4N$ contributions for the maximum values of $A_y$ in $\nH$ 
scattering which is closely related to $\pHe$ $A_y$.
The effects of the $2N$ dispersion and effective $3N$ force are quite
sizable, about -15\% and 10\%, respectively,
but partially cancel each other. A similar canceling was observed
in the $3N$  $A_y$ \cite{deltuva:03a} though there the $3N$ force effect
was larger than the $2N$ dispersion, in contrast to the $4N$ system.

\begin{figure}[!]
\begin{center}
\includegraphics[scale=0.48]{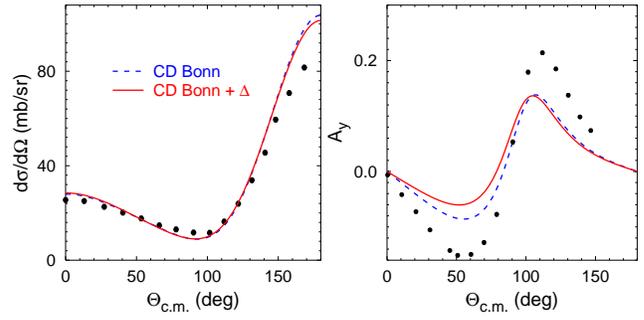}
\end{center}
\caption{\label{fig:pt-nh} 
Differential cross section and proton analyzing power of the
$p+\Hh \to n + \He$ reaction at 6 MeV proton lab energy.
Curves as in \Fig~\ref{fig:nt}.
The cross section data  are   from \Ref~\cite{wilson:61}. 
$A_y$ data are  from \Ref~\cite{jarmer:74}.}
\end{figure}

\begin{figure}[!]
\begin{center}
\includegraphics[scale=0.48]{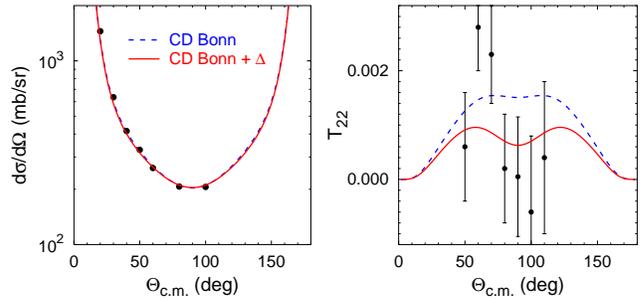}
\end{center}
\caption{\label{fig:dd} 
Differential cross section and deuteron tensor analyzing power $T_{22}$
of the elastic $\dd$ scattering at 3 MeV deuteron lab energy.
Curves as in \Fig~\ref{fig:nt}.
The cross section data  are from \Ref~\cite{blair:48a} and $T_{22}$ data
are from \Ref~\cite{crowe:00a}.}
\end{figure}

The elastic differential cross section in the coupled $\pH$ and $\nHe$ reactions
correlates to some extent with that of $\pHe$ and $\nH$ scattering and is
similarly reduced by the $\Delta$ excitation at forward and backward angles.
In contrast, the effect is much weaker for the $p+\Hh \to n + \He$
transfer cross section as shown in Fig.~\ref{fig:pt-nh}. The $\Delta$ effect
on $A_y$ in this reaction is consistent with the findings of 
\Ref~\cite{deltuva:07c} where increasing the $3N$ binding energy
moves the predictions away from the data. 
The $\Delta$ effect is tiny for the elastic $\dd$ cross section
as shown in Fig.~\ref{fig:dd}, but is visible
for the deuteron tensor analyzing powers which, however, are very small.

\begin{figure}[!]
\begin{center}
\includegraphics[scale=0.55]{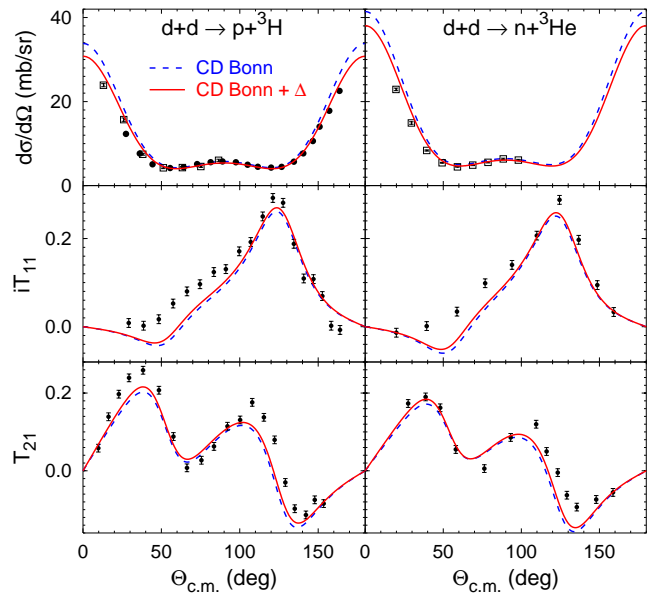}
\end{center}
\caption{\label{fig:dd-ptnh} 
Differential cross section and deuteron analyzing powers of the
$d+d \to p+\Hh$ and  $d+d \to n+\He$ reactions at 3 MeV deuteron lab energy.
Curves as in \Fig~\ref{fig:nt}.
The cross section data  are from \Refs~\cite{blair:48b} (squares)
and \cite{gruebler:72a} (circles). Analyzing power data are from
\Ref~\cite{gruebler:72a} for $d+d \to p+\Hh$ and from
\Ref~\cite{dries:79a} for $d+d \to n+\He$.}
\end{figure}

As we found in \Ref~\cite{deltuva:07c} the observables of the two 
charge-symmetric transfer reactions $d+d \to p+\Hh$ and  $d+d \to n+\He$ 
correlate to some extent with the $3N$ binding energy
and with the deuteron $D$-state probability.
That former correlation is reflected in Fig.~\ref{fig:dd-ptnh} where the
inclusion of the $\Delta$-isobar excitation brings the theoretical
predictions closer to the data for the differential cross section,
but has a smaller effect on the analyzing powers.

\section{Conclusions \label{sec:concl}}

The technical apparatus, developed in 
\Refs~\cite{deltuva:07a,deltuva:07b,deltuva:07c} for the solution of the
$4N$ bound state and scattering equations, is employed and extended.
The extension covers the use of a realistic coupled-channel potential
allowing for the excitation of a single nucleon to a
$\Delta$ isobar. The $\Delta$ isobar mediates effective $2N$, $3N$ and $4N$
forces, consistent with each other. A procedure for isolating the
$\Delta$-isobar effects of $2N$, $3N$ and $4N$ nature on observables
is given and used to study different dynamic mechanisms.

\emph{Technically}, this paper demonstrates that $4N$ calculations
with realistic $2N$, $3N$ and $4N$ forces are feasible. The Coulomb
repulsion between protons is included. Fully converged results
for the $4N$ binding energy and for the $4N$ scattering observables
below three-body breakup threshold are obtained.

\emph{Physicswise}, this paper shows for the first time that, within the present 
model space, $4N$ force effect on nuclear observables is much smaller than the 
$3N$ force effect. That fact is shown in the framework of $\Delta$-mediated
effective many-body forces for the $4N$ bound state and low-energy scattering 
observables, and is a very valuable confirmation of the traditional belief 
in a hierarchic order for the importance of many-nucleon forces.
However, the inclusion of the 
$\Delta$ isobar is unable to resolve the long-standing discrepancies
with the experimental data, neither for the $\nH$ total cross section 
in the resonance region nor for $A_y$ in $\pHe$ scattering.  Nevertheless,
in addition to possible differences resulting from model dependence, some
differences between the work of Lazauskas \& Carbonell~\cite{lazauskas:04a}, 
the Pisa group~\cite{viviani:01a,fisher:06} and the present calculations need 
to be sorted out in the near future. While our findings seem to coincide with 
those of the Pisa group for $\pHe$ vis-a-vis the effect of the $3N$ force, 
the calculations by Lazauskas \& Carbonell indicate that 
the Urbana IX $3N$ force, when added to AV18 $2N$ force, bears almost no effect
on the total cross section $\sigma_t$ in the $\nH$ resonance region. Given that 
$\nH$ and $\pHe$ only differ by the Coulomb interaction and small charge dependent  
terms in the $2N$ force, one does not expect such a different behavior between 
$\nH$ and $\pHe$ when $3N$ forces are added.

\emph{Dynamically}, our calculations are based on the two-baryon 
coupled-channel potential CD Bonn + $\Delta$~\cite{deltuva:03c}, which fits the 
deuteron properties and $2N$ elastic scattering data as well as the best $2N$ 
potentials~\cite{wiringa:95a,machleidt:01a,doleschall:04a,entem:03a}. 
As standard in the description of nuclear structure and scattering,
the $\Delta$ isobar is assumed to be a stable particle of fixed mass
without subthreshold corrections arising from the $\Delta$-generated
$P_{33}$ $\pi N$ resonance. This assumption is a crude and in principle
unnecessary simplification of the $\Delta$ isobar's dynamic structure,
not allowing a direct application of the coupled-channel potential to
pionic reactions. This fact is the reason why the $\Delta N$ and especially
the $\Delta \Delta$ parts of the two-baryon potential in its present form
are not sufficiently constrained by the $2N$ data; their full determination
requires the data of $\pi NN$ dynamics.
We recall that all nuclear potentials are indeterminate to some extent, e.g., $2N$ 
potentials with respect to their short-range behavior and their amount of 
nonlocality. However we are especially concerned about the indeterminacy 
of the  $\Delta \Delta$ part of 
the employed two-baryon coupled-channel potential, since it is responsible for the 
higher-order $3N$ force and for the $4N$ force, the focus of this paper. Fortunately, 
that dynamic indeterminacy does not change our physics conclusion in any form: 
Exploiting that model dependence of the potential by using, besides CD Bonn + $\Delta$,
also its alternative CD Bonn + $\Delta'$ described in the Appendix, the $4N$ force 
effect on binding energy remains much smaller than the $3N$ force effect
and for the scattering observables it is completely negligible. Finally, completing the 
list of all possible shortcomings of the employed two-baryon coupled-channel potential, 
the $2N$ potential in the presence of a $\Delta$ isobar, encountered in the Hamiltonian 
underlying our calculations, is constrained only by the data of $\pi NNN$ dynamics;
however, different choices for that part of the potential appear inconsequential for 
all studied observables  according to our findings in the $3N$ system.

\emph{In a longer range vision}, the extension of the Hamiltonian of this paper to 
cover also pionic reactions is quite possible,
pushing the descriptions of $3N$ and $4N$ scattering to intermediate energies and
thereby decreasing the model dependence inherent in the chosen force model with 
$\Delta$-isobar excitation. Furthermore, we hope for a consistent derivation and 
tuning of purely nucleonic $2N$, $3N$ and $4N$ forces in the framework of chiral EFT; 
their subsequent application to the $4N$ observables, studied in this paper, would be 
a challenging enterprise and a wonderful alternative to our present work.

We thank R.~Lazauskas and M.~Viviani for the discussion of $3N$ force effects.
A.D. is supported by the Funda\c{c}\~{a}o para a Ci\^{e}ncia e a Tecnologia
(FCT) grant SFRH/BPD/34628/2007
and A.C.F. in part by the FCT grant POCTI/ISFL/2/275.


\begin{appendix}
\section{Alternative two-baryon coupled-channel potential CD~Bonn~+~$\Delta'$} 

This paper works predominantly with the two-baryon coupled-channel potential 
CD Bonn + $\Delta$, derived in \Ref~\cite{deltuva:03c}.
Its particular feature is a strong $\sigma$-meson coupling 
$g_{\sigma NN} g_{\sigma \Delta\Delta}/4\pi = 8.7$ 
for the direct $v_{\Delta\Delta}$ component in $N\Delta$ states ${}^5SDG_2$
coupled to the nucleonic partial wave ${}^1D_2$.
That coupling is undetermined within sizable
limits; it was chosen to obtain more binding in the $3N$ bound state,
nevertheless allowing for the optimal fit of $\chi^2/\mathrm{datum}=1.02$
to the elastic $2N$ scattering data. This appendix presents selected results
for an alternative coupled-channel potential CD Bonn + $\Delta'$
\cite{deltuva:04b} with weaker $g_{\sigma NN} g_{\sigma \Delta\Delta}/4\pi = 5.0$
that, after refitting the parameters of the two $\sigma$ mesons in the
nucleonic part $v_{NN}$ of the potential,
 still allows for an optimal description of $2N$ data with
$\chi^2/\mathrm{datum}=1.02$. For completeness in Table~\ref{tab:cdbdx}
we give the changed $\sigma$-meson parameters of the CD Bonn + $\Delta'$ 
potential in the ${}^1D_2$ partial wave; other parameters have the same
values as for CD Bonn + $\Delta$ and are given in \Ref~\cite{deltuva:03c}.

\begin{table}[h]
\begin{tabular}{l*{4}{@{\quad\quad}r}}
\hline
 & $ m_{\sigma_i}$ &  $g_{\sigma_i}^2/4\pi \,(pp)$  & 
$g_{\sigma_i}^2/4\pi \,(np)$  &  $g_{\sigma_i}^2/4\pi \,(nn)$ 
\\ \hline
${\sigma_1}$ & 350 & 0.50683 & 0.51269 & 0.51424 \\ 
${\sigma_2}$ & 1225 & 148.10  & 148.42  & 149.28 \\
\hline
\end{tabular}
\caption{ \label{tab:cdbdx}
$\sigma$-meson parameters for the potential CD Bonn + $\Delta'$
in the nucleonic partial wave ${}^1D_2$. The masses $ m_{\sigma_i}$ are in MeV.}
\end{table}

Table \ref{tab:ebx} shows the changes in the binding energy of $\Hh$, $\He$
and $\HE$ arising for CD Bonn + $\Delta'$.
The $\Delta$-isobar effect is weaker than for CD Bonn + $\Delta$, turning
even non-beneficial for $\HE$. Whereas the $2N$ dispersive and $3N$
Fujita-Miyazawa effects, i.e., $\Delta E_2$ and $\Delta E_{3}^{\mathrm{FM}}$,
remain practically unchanged, the higher order $3N$ and $4N$ force
contributions, i.e., $\Delta E_3^{\mathrm{h.o.}}$ and $\Delta E_{4}$,
being much more sensitive to the $v_{\Delta\Delta}$ component, get strongly
reduced, thereby changing the complete $\Delta$-isobar effect considerably. 
This is a measure of our model dependence for 0.17 $> \Delta E_{4} >$ 0.03.

\begin{table}[h]
\begin{tabular}{l*{3}{@{\quad\quad}r}}
\hline
& ${}^3\mathrm{H}$ & ${}^3\mathrm{He}$ &  ${}^4\mathrm{He}$ \\ \hline
CD Bonn  & 8.00 & 7.26 & 26.18  \\
CD Bonn + $\Delta'$ &    8.05 &  7.31 &  25.89 \\
exp &   8.48 & 7.72 & 28.30  \\ \hline
 $\Delta E_2$ & -0.51 & -0.48  & -2.78  \\
$\Delta E_{3}^{\mathrm{FM}}$ & 0.50 & 0.48 & 2.20 \\
$\Delta E_3^{\mathrm{h.o.}}$ & 0.06 & 0.05 & 0.26 \\
$\Delta E_{4}$ & & & 0.03 \\
\hline
\end{tabular}
\caption{ \label{tab:ebx}
Same as Table~\ref{tab:eb}, but with CD Bonn + $\Delta'$.}
\end{table}

\end{appendix}

\end{document}